\documentclass[12pt]{article}

\usepackage{graphicx}
\usepackage{rotating}
\usepackage{amssymb}

\usepackage[cp1251]{inputenc}
\usepackage[greek,ukrainian,english]{babel}

\usepackage{amsfonts}

\def\NkoYelen{\raisebox{-0.5ex}{\includegraphics[clip,scale=1.00]{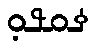}}}

\def\be{\begin{eqnarray}}
\def\ee{\end{eqnarray}}
\def\ben{\begin{eqnarray*}}
\def\een{\end{eqnarray*}}

\def\eps{\varepsilon}

\begin{document}

\title{Application of a Quantum Ensemble Model to Linguistic Analysis}

\author{Andrij Rovenchak$^1$\footnote{%
Corresponding author: A. Rovenchak, Department for Theoretical Physics, 12 Drahomanov St., Lviv, UA-79005, Ukraine;
tel.: +380 32 2614443, e-mail: {\tt andrij.rovenchak@gmail.com, andrij@ktf.franko.lviv.ua}
}\ \ and Solomija Buk$^2$\\
$^1$\,Department for Theoretical Physics,\\
$^2$\,Department for General Linguistics,\\
Ivan Franko National University of Lviv, Ukraine\\
}

\maketitle

\begin{abstract}
A new set of parameters to describe the word frequency behavior of texts is proposed. The analogy between the word frequency  distribution and the Bose-distribution is suggested and the notion of ``temperature'' is introduced for this case. The calculations are made for English, Ukrainian, and the Guinean Maninka languages. The correlation between in-deep language structure (the level of analyticity) and the defined parameters is shown to exist.


\end{abstract}

\newpage
\section{Introduction}
Quantitative analysis of large text samples revealed regularities in the behavior of various text parameters.
The empirical laws found in texts, such as Zipf's law, are known to hold in various domains,
in particular the distribution of nucleotides in genomes and other fields of biology \cite{gen1,gen2,Dudek,Ogasawara}, regularities in social sciences \cite{Urzua,soc1,soc2,soc3,soc4}, etc.

Approaches from the domain of statistical physics can be used to study systems composed of many units in general, and texts are suitable for such studies as well.
The application of physical techniques in linguistic is quite common \cite{Kanter,Fontnari,Ferrer-i-Cancho,Kechedzhi,Bernhardsson}, other domains are also successfully covered by physical approaches, cf.~\cite{Holovatch}.

In this work, we analyze quantitative behavior of texts by finding analogy with a bosonic system within grand canonical ensemble. In doing so, we demonstrate the possibility to assign some new parameters characterizing the frequency structure of texts, one of which can be conventionally called ``temperature''.

The notion of ``temperature of texts'' was discussed from different points of view by several authors. Mandelbrot \cite{Mandelbrot} suggested the name ``informational temperature of texts'' for a parameter in a rank--frequency distribution (known as the Zipf--Mandelbrot law). Such a parameter is related to ``good'' or ``bad'' employment of words, especially rare words \cite{poetics}. The ``temperature'' as a measure of communicative ability was introduced in \cite{Kosmidis}.
 Recently, Miyazima and Yamamoto \cite{Miyazima} used the classical Boltzmann distribution to define the ``temperature of texts'' from the frequency data of the most frequent words. We propose a different approach, mainly addressing the behavior of low-frequency vocabulary.

The paper is organized as follows. In Section~\ref{sec:RFD} we recall main notions used in further text, namely the principles of rank--frequency distribution compilation as well as the term \textit{hapax legomena}. Section~\ref{sec:PhysicalAnalogy} contains main part, where the physical analogy with the Bose-distribution is discussed in detail and parameters of text frequency distribution are given suitable interpretation. The results of text analysis in three languages are given in Section~\ref{sec:Results}, and Section~\ref{sec:Discussion} contains brief discussion of the presented approach.

\section{Rank--frequency distribution}\label{sec:RFD}

In this work, we analyze texts on the word level. While the notion of ``word'' has no unique definition, cf.~\cite{Popescu}, we restrict ourselves to the so called ``orthographic word'' defined as an alphanumeric sequence between two spaces or punctuation marks. Different word forms, like `hand' and `hands', `write' and `wrote', etc. are considered as different words for simplicity.

To obtain a rank--frequency distribution, one should first compile the frequency list from a given sample. Then, the item with the highest frequency is given {\bf rank 1}, the second most frequent item is given {\bf rank 2}, and so on.
The items with the same frequency are given a consecutive range of ranks, the ordering within which can be arbitrary.

The studies of rank--frequency distributions originate from text analysis, and despite the regularities found there are known to hold in various domains, texts still remain the most easily accessible material having a good variety of sorts to be analyzed.

\begin{figure}[ht]
\centerline{\includegraphics[clip,scale=0.75]{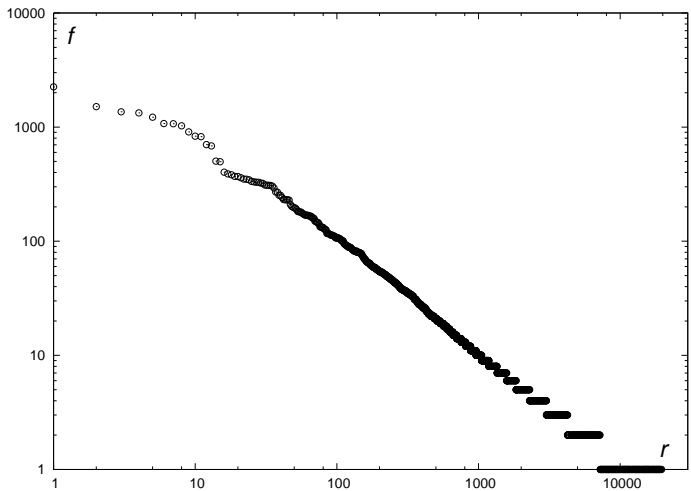}}
\caption{Typical rank--frequency distribution. The absolute frequency $f$ is shown versus the rank $r$ for orthographic words of
\textit{Perekhresni ste\v{z}ky [The Cross-Paths]}, a Ukrainian novel by Ivan Franko. Data are obtained by the authors on the preliminary stage of
compiling the frequency dictionary of the novel \cite{PS-freqdict}.}
\label{fig:typicalRFD}
\end{figure}

A typical rank--frequency distribution has the shape shown in Fig.~\ref{fig:typicalRFD}.

\bigskip

Horizontal plateaus in the domain of high ranks / low frequencies correspond to a large number of words having the same frequency.
The longest plateau correspond to frequency 1. Such words are known as \textit{hapax legomena}, the term originating from Bible studies.

\textit{Hapax legomena} is a Plural of the Classical Greek term \textit{hapax legomenon}
({\greektext <'apax leg'omenon})
translated as `[something] said [only] once'.
That is, this term corresponds to the tokens appearing only once in a given sample.
Examples from Bible include \cite{JE-hapax}:
\raisebox{-0.2ex}{\includegraphics[clip,scale=1.00]{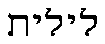}}
`Lilith' (a word of obscure meaning) or
\raisebox{-0.7ex}{\includegraphics[clip,scale=1.00]{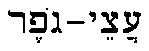}}
 `gopher wood' (used to build Noah's Ark).
Other often cited examples are:
{\it\greektext a>ut'oguon}, a kind of plough (Hesiod, {\it\greektext >'Erga ka`i <Hm'erai} \textit{[Opera et Dies = Works and Days]}, 433); 
{\it honorificabilitudinitatibus} \ `the state of being able to achieve honors' (Shakespeare,
\textit{Loves Labours Lost}, act 5, scene 1 \cite[p.~372]{Shakespeare}).

For large text samples, about 40 to 60 per cent of occurring words are hapaxes, depending on the text size \cite[p.~72]{Kornai}.
The relative number of \textit{hapax legomena} slightly decreases as the text becomes longer. Various quantities depending on the text size $N$ are well described by the power law \cite{Tuldava1987}, and the number of hapaxes fits into this family as well,
$$
N_{\rm hapax} \stackrel{?}{=} A N^b.
$$

Note, however, that for statistical studies texts must be sufficiently long.

\textbf{\textit{Indeed, even in such a long sentence having twenty-three tokens all the words are \underline{hapaxes}, except for ``\underline{hapaxes}'' themselves since they occur twice.}}

\section{Physical analogy}\label{sec:PhysicalAnalogy}
The rank--frequency distribution of words in texts has clear similarities with Bose-distribution in statistical physics. We suggest to identify
the energy level numbers $j$ with word frequencies (the number of occurrences in a given text). Thus, the words with frequency 1 occupy the level $j=1$,
the words with frequency 2 occupy the level $j=2$, etc.
The level occupation then corresponds to the number of different words with the same frequency.
Since the level occupation can reach any value (in particular, significantly larger than unity) the use of the Bose-distribution is appropriate. The lowest level corresponds to \textit{hapax legomena} and in this scheme can be identified with the Bose-condensate.

\subsection{Defining energy spectrum}
In the Bose-distribution the occupation of the $j$th level is given by
\be\label{eq:Nj}
N_j = {1\over z^{-1}e^{\eps_j/T}-1},
\ee
where $z$ is the fugacity, $\eps_j$ is the energy of the $j$th level, and $T$ is the temperature.

As shown further, a power energy spectrum gives a proper description for lower levels,
\be\label{eq:spectrum}
\eps_j = (j-1)^\alpha.
\ee
The unity is subtracted to ensure that the lowermost level has zero energy.

Due to the nature of the frequency distribution, a simple model of a very weak log-of-log growth is appropriate for the energy spectrum at high levels,
$\varepsilon_j \propto \ln\ln j$ for $j\gg1$, cf.~Fig.~\ref{fig:energy-fit}.
Note, however, that a log-of-log spectrum requires the maximal number of levels to be bounded from above by some $j_{\max}$.

\subsection{Parameters of the Bose-distribution}
We defined the parameters in Eq.~(\ref{eq:Nj}) in two steps. First, the parameter $z$, being interpreted as fugacity in physics, is defined from the occupation number of the lowermost state, i.\,e., the number of \textit{hapax legomena}:
\be\label{eq:Nhapax}
N_{\rm hapax} = {z \over 1-z}.
\ee

\begin{figure}[ht]
\centerline{\includegraphics[clip,scale=0.75]{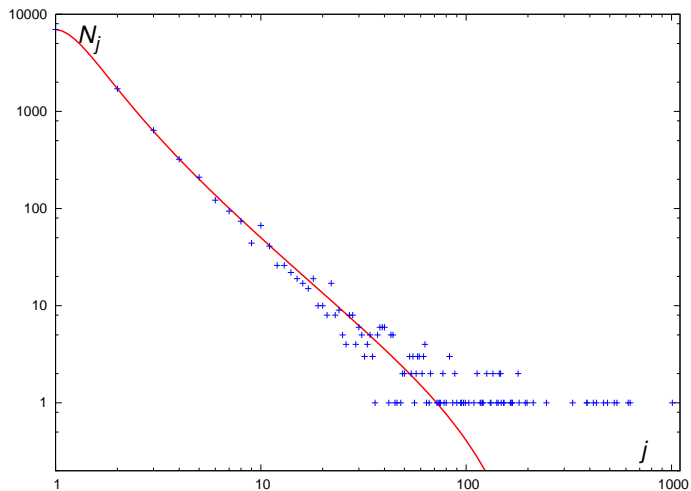}}
\caption{(Color online) The fit of the power energy spectrum to the level occupations.
Blue crosses correspond to the data obtained by the authors on the basis of first 40 chapters (of total 60) from the text mentioned
in the caption of Fig.~\ref{fig:typicalRFD}. Solid line is the fitting curve (\ref{fitting_curve}) for the first 20 values of occupation numbers $N_j$.}
\label{fig:energy-fit}
\end{figure}

``Temperature'' $T$ and exponent $\alpha$ in Eq.~(\ref{eq:spectrum}) are found simultaneously by fitting the occupation of higher energy levels to
\be\label{fitting_curve}
N_j = {1\over z^{-1}e^{(j-1)^\alpha/T}-1}
\ee
via two parameters, $\alpha$ and $T$. The sample results of fitting are presented in Fig.~\ref{fig:energy-fit}.
These calculations, as well as other given further in these work, were made using the nonlinear least-squares Marquardt--Levenberg algorithm implemented in the
{\tt fit} procedure of GnuPlot, version 4.0.

One should note that the parameter $T$ is dimensionless in our case, as is the energy $\eps_j$.
Such definition differs, e.\,g., from \cite{Miyazima}, where a distribution of some standard text was used to set the reference temperature in Kelvins.

The state with $T=0$ corresponds to all the frequencies equal to unity, that is, the whole text is composed of \textit{hapax legomena}. This could be the case of a very short text, not longer than just one or a couple of sentences (cf.\ the example at the end of Section~\ref{sec:RFD}).

Presently, we fit first 10--20 levels using the power excitation spectrum (\ref{eq:spectrum}). Higher levels are neglected since a different dependence on $j$
must be applied to ensure good fitting of the occupation data $N_j$, a suggested in the previous subsection.
The parameter $T$ obtained in such a way scales (very precisely) as $N^\beta$ ($\beta<1$).
The scaling is related to the definition of ``thermodynamic limit'' for the problem under consideration. Just to recall, in the system of $N$ bosons trapped to a $D$-dimensional harmonic oscillator potential with frequency $\omega$ the thermodynamic limit is given by
$\omega N^{1/D}=\rm const$ as $N\to \infty, \omega\to0$ \cite{Posazhennikova}.
Since $\omega$ (or $\hbar\omega$ if Planck's constant $\hbar$ is not set equal to unity) is a natural unit for the oscillator energy,
the power-like scaling of the quantities measured in the energy units is expectable for the and for the systems with power energy spectrum as well.

Curiously, the ratio $\ln T / \ln N$ exhibits an insignificant variation with the size of the text sample (for a sufficiently long text). This makes it a good variable for comparative linguistic studies.

\section{Some results}\label{sec:Results}

So far, we have performed analysis of some texts written in English (Germanic language), Ukrainian (Slavic language), as well as Guinean Maninka (in the Nko script; a language from the Mande family). Such a vast choice is suggested to check the approach on significantly different language materials in order to reveal both universal and unique features of the parameter behavior.

\begin{figure}[ht]
\centerline{\includegraphics[clip,scale=0.75]{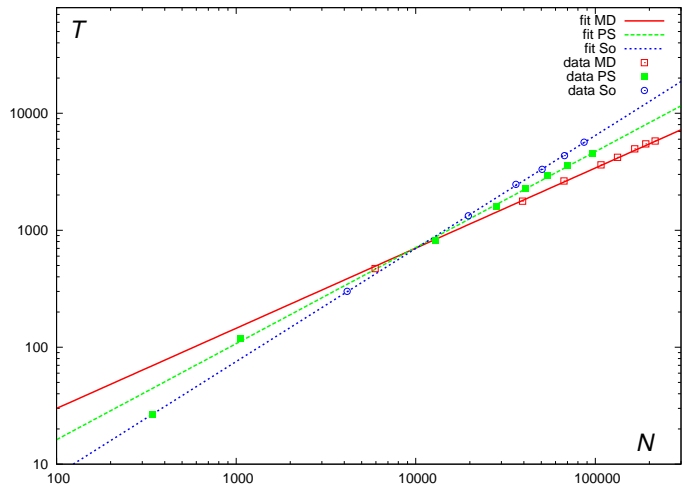}}
\caption{(Color online) The behavior of ``temperature'' as the size of text grows.
MD --- \textit{Moby-Dick}; PS --- \textit{Pe\-re\-khres\-ni ste\v{z}ky}; So --- \textit{Sobor}).
The lines correspond to the linear fits of the data represented by the respective symbols.
}
\label{fig:temp-scale}
\end{figure}

Fig.~\ref{fig:temp-scale} demonstrates the ``temperature'' behavior of an English text (\textit{Moby-Dick} by Hermann Melville) and two novels in Ukrainian (\textit{Pe\-re\-khres\-ni ste\v{z}ky [The Cross-Paths]} by Ivan Franko and \textit{Sobor [The Cathedral]} by Oles Honchar).

\medskip
Table~\ref{tab:results} shows the numerical data on the parameter $T$ calculated by grasping increasing shares of chapters.
The data based on an article from a Guinean journal \textit{Y\'el\'en$\!$\`{}} are also given.

The values of $z$ in all the cases are close to 1. A better resolution might be achieved by introducing an analog of the chemical potential $\mu$ in a standard way, $z=e^{\mu/T}$.

The fitting gives the values of the exponent $\alpha$ slightly decreasing as the text size grows.
An interpretation in terms of an external potential can be applied to justify such a change.
Indeed, if the presence of an external potential is treated in the semiclassical approach \cite{Bagnato}, the decreasing values of the exponent in a power excitation spectrum effectively correspond to weakening of the steepness of an external potential. That is, as a text becomes longer, it suffers less from some external influences.

Indeed, in one dimension a power energy spectrum $\eps_p\propto p^{\alpha}$ leads to the density of states
\be
g(\eps)\propto\eps^{{1\over\alpha}-1}.
\ee
On the other hand, non-interacting particles confined into trapping potential
$U(x)\propto x^\eta$ in the semi-classical approach \cite{Bagnato} have the density of states
\be
g(\eps)\propto \eps^{{1\over \eta}-{1\over2} }.
\ee
Note that a rigid box corresponds to $\eta=\infty$.

Thus, an effectively occurring exponent $\alpha$ is related to $\eta$ via
\be
\alpha = {2\eta\over \eta+2}
\ee
leading to $\alpha=3/2$ for $\eta=6$ and $\alpha=1$ for $\eta=2$.

Preliminary, the obtained $T$ and $\alpha$-exponent values correlate with the analyticity level of the language. Lower values correspond to higher analyticity (less word inflection), as can be seen from the opposition between English and Ukrainian (both Indo-European languages). So far, we do not have sufficient data to make further statements, in particular, for the language from an unrelated language family (Mande), which data are given for curiosity and future references. As should be expected, a low value of $\alpha$ for the Maninka sample suggests a high level of analiticity.

\begin{figure}[ht]
\centerline{\includegraphics[clip,angle=-90,scale=0.5]{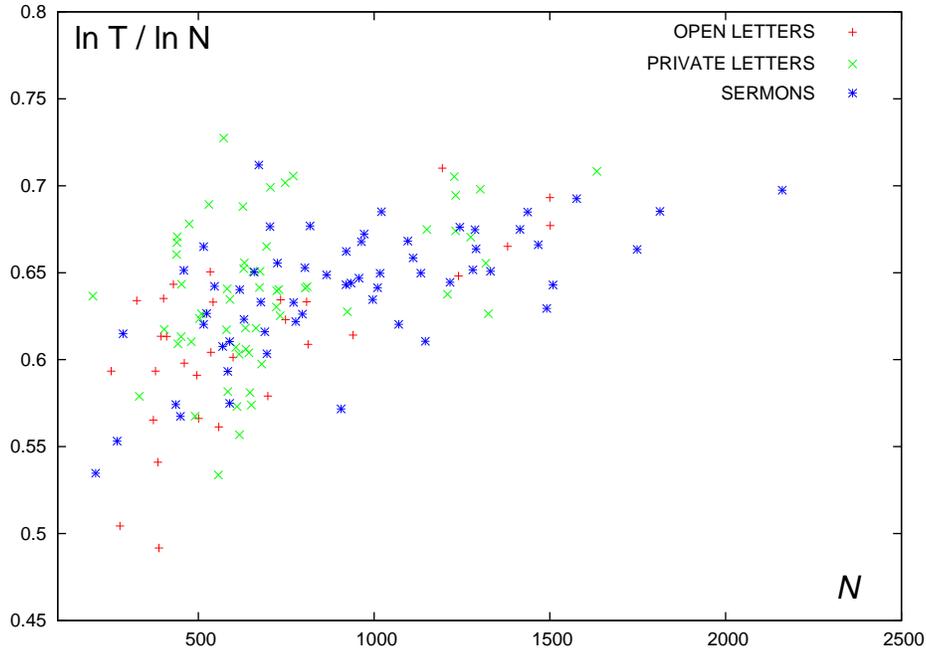}} 
\caption{(Color online) The behavior of $\ln T/\ln N$ for texts from different genres. Open letters, private letters, and sermons are shown.}
\label{fig:lnTlnN}
\end{figure}

Finally, in Fig.~\ref{fig:lnTlnN} we present the results of ``temperature'' calculation made for short Ukrainian texts of different genres \cite{project}. Close values denote weak genre dependence of this parameter. A multivariate discriminant analysis is required to study this issue in more detail.

\section{Brief discussion}\label{sec:Discussion}

The presented results are a preliminary attempt to define a new set of parameters describing frequency structure of texts.
Further application of this approach to a larger number of texts from different languages written by different authors is required to establish the correlation
between parameter values and text language/authorship.
The shape of the spectrum in the whole domain of the variation of $j$ must be considered in further studies to give a proper description of the level occupation.
Also, more parameters can be calculated within the ``thermodynamic approach''
(like some analogs of total energy, specific heat, etc., cf.~\cite{Kosmidis}). One of the tasks which we expect from such calculations is the possibility of automatic text attribution useful for automated language processing. Applications beyond linguistics -- in genetics, social sciences, etc. -- are also possible in future.

\section*{Acknowledgements}
We are grateful to anonymous Referees for useful critical comments and suggestions, which were helpful in improving the manuscript.

This work was partly supported by Project No. M/6-2009 from the Ministry of Education and Sciences of Ukraine and
WTZ Project UA 05/2009 from the \"Osterreichischer Austauschdienst.

\clearpage

\listoffigures


\begin{table}[ht]
\caption{The parameters of ``energy spectrum'' and ``temperature'' of texts}\label{tab:results}
\centerline{
\begin{tabular}{rcrcc}
\hline
$N$\ \  &   $\alpha$    &   $T$ \ \ &   $
{\ln T/
\ln N}$ &   $T/N$   \\
\hline
\multicolumn{5}{l}{\textit{Moby-Dick} (ENG)}\\
\hline
5942    &   1.97    &   470.4   &   0.708   &   0.0792  \\
39363   &   1.60    &   1773.3  &   0.707   &   0.0451  \\
66916   &   1.56    &   2639.7  &   0.709   &   0.0394  \\
107503  &   1.48    &   3622.3  &   0.707   &   0.0337  \\
132968  &   1.48    &   4207.3  &   0.707   &   0.0316  \\
165746  &   1.48    &   4968.4  &   0.708   &   0.0300  \\
191040  &   1.47    &   5476.5  &   0.708   &   0.0287  \\
215270  &   1.45    &   5791.3  &   0.706   &   0.0269  \\
\hline
\multicolumn{5}{l}{\textit{{\selectlanguage{ukrainian}Перехресні стежки}
(The Cross-Paths)
} (UKR)}\\
\hline
343 &   1.57    &   26.6    &   0.562   &           0.0774  \\
1052    &   2.03    &   119.1   &   0.687   &           0.1132  \\
12949   &   1.68    &   812.0   &   0.708   &           0.0627  \\
28010   &   1.73    &   1610.1  &   0.721   &           0.0575  \\
40811   &   1.72    &   2270.7  &   0.728   &           0.0556  \\
54361   &   1.70    &   2964.3  &   0.733   &           0.0545  \\
70330   &   1.64    &   3597.4  &   0.734   &           0.0512  \\
96083   &   1.57    &   4561.4  &   0.734   &           0.0475  \\
\hline
\multicolumn{5}{l}{\NkoYelen \ \textit{Y\'el\'en$\!$\`{} (The Light)} journal (NKO)}\\
\hline
429   & 1.42 &   45.2  & 0.629  &           0.1053 \\
\hline
\end{tabular}
}
\end{table}
\end{document}